\documentclass[preprint,aps,12pt,preprintnumbers,eqsecnum,nofootinbib]{revtex4}
\usepackage{graphicx}
\usepackage{subfigure}

\usepackage{color}
\usepackage{amssymb,amsmath}
\begin{document}
%
%
\title{\vspace*{0.5in} 
An Introduction to Holographic QCD for Nonspecialists\footnote{Invited review for {\em Contemporary Physics.}}
\vskip 0.1in}
\author{Joshua Erlich\footnote{jxerli@wm.edu}}

\affiliation{High Energy Theory Group, Department of Physics,
College of William and Mary, Williamsburg, VA 23187-8795}
\date{\today}
\begin{abstract}
Holographic QCD is an extra-dimensional approach to modeling hadrons, the bound states of the strong interactions.  In holographic models, the extra spatial dimension  creates a waveguide for fields, and the discrete towers of modes  propagating in that waveguide are interpreted as hadronic resonances.  These models are motivated by the AdS/CFT correspondence, which is a duality  that relates theories  in different numbers of spatial dimensions.   Holographic models have the potential to provide  a better understanding of    strongly interacting systems of quarks and gluons,  as well as  unconventional superconductors and other nonperturbative systems.  

\end{abstract}
\maketitle

\section{Introduction}
In March 1984, the cover of National Geographic magazine featured a holographic image of an eagle  \cite{NatGeo}.  This was the first time that a hologram was  mass-circulated in a major publication, and the magazine helped to develop a widespread enthusiasm for  holography, a technology which was already two decades old at that time \cite{Leith-Upatnieks}.  The hologram gives the illusion of a three-dimensional image from a two-dimensional surface, created by diffraction of light from the surface.  However, closer inspection of a holographic image reveals that details of the higher-dimensional object are lost, and the hologram reproduces only an approximation of the image of the original object.  

It would be wonderful if a perfect reproduction of a higher-dimensional object could be created from information stored in a lower-dimensional one, but intuition demands that lower-dimensional systems do not have enough degrees of freedom to  completely describe  higher-dimensional systems.  Yet, string theorists have discovered that something analogous to a perfect holographic reproduction is, in fact,  possible: physical theories in different numbers of spatial dimensions can  contain identical information, so that calculations of observables in a lower-dimensional version of the theory may be translated into  calculations in the higher-dimensional version of the theory, and {\em vice versa}.  Situations  in which there is more than one way to describe the same physics are called {\em dualities}, and if the different descriptions are also in different numbers of spatial dimensions then they are called {\em holographic dualities}.

The first example of a holographic duality  was discovered by Juan Maldacena in 1997 \cite{Maldacena:1997re} in the context of string theory, and that discovery created a new field of research which has received constant attention from the string theory community in the over  decade and a half since.  
Maldacena's  conjectured duality, known as the Anti-de Sitter/Conformal Field Theory (AdS/CFT) correspondence, was formal in its nature and does not appear to be directly related to the real world.  However, many additional examples of holographic dualities are now known, some of which bear intriguing similarities to real-world physical systems.  A remarkable feature of holographic dualities is that in many cases one description of the physics is strongly coupled, while the other description is  weakly coupled and amenable to perturbative analysis of observables.  
As a result, the existence of holographic dualities has motivated a novel way to model complex nonperturbative physical systems, including  the strong nuclear interactions \cite{Kruczenski:2003uq,Babington:2003vm,Sakai-Sugimoto, deTeramond:2005su,Erlich:2005qh,Da Rold:2005zs}, technicolor-like models of electroweak symmetry breaking \cite{holoEWSB},  unconventional superconductors \cite{holosup} and other condensed matter systems \cite{Green}.  This brief review for nonspecialists will focus on the application to the strong interactions.

\section{The Strong Interactions and QCD}
The strong interactions are described in the Standard Model by a quantum field theory known as quantum chromodynamics (QCD).
The strong interactions are responsible for binding quarks into protons, neutrons, and other hadrons.  They are responsible for the existence of bound nuclei inside atoms and for trapping the energy which is released during the fission reactions that power some cities.  Without the strong interactions there could be no  atoms, and the universe would be flooded with fractionally charged quarks.  The strong interactions are an integral part of the Standard Model of particle physics, yet it is notoriously difficult to make precise predictions about observables that are sensitive to the strong interactions, such as the masses and decay rates of the hadrons.  

One basic difficulty is that the theory of the strong interactions, Quantum Chromodynamics (QCD), does not contain a small dimensionless quantity  that would allow for perturbative calculation of low-energy observables.  In contrast, the quantum theory of electromagnetism, Quantum Electrodynamics (QED), contains a small parameter which determines the relative size of subsequent terms in perturbative calculations, namely the low-energy fine structure constant $\alpha\approx1/137$.  The small magnitude of $\alpha$ is what makes precise perturbative  calculations of electromagnetic observables like the spectrum of the hydrogen atom and the cross section for Compton scattering feasible.  Typical  courses on quantum field theory include a perturbative calculation of the gyromagnetic ratio of the electron (the ratio of the electron's magnetic moment to its spin angular momentum), sometimes as a homework assignment, to six significant figures.  On the other hand, perturbative calculations of low-energy hadronic observables from QCD are not possible, so one of two approaches is used to proceed.  By using clever computational techniques on high-powered computer clusters, lattice QCD simulations are able to determine certain hadronic observables ({\em e.g.} light hadron masses \cite{Fodor:2012gf}) by defining QCD on a lattice of spacetime points, and then extrapolating to a continuum limit. The alternative approach is to model the strong interactions based on some of its important features, such as its exact and approximate symmetries,  supplemented with bits of experimental data which then allow for estimates of additional observables.  For example, chiral perturbation theory makes use of a symmetry that QCD would have if the up and down quarks were massless in order to model phenomena involving bound states of those quarks.

The comparison of Quantum Chromodynamics and Quantum Electrodynamics is instructive, as the fundamental descriptions of the two theories are in some ways quite similar.  Both Quantum Chromodynamics and Quantum Electrodynamics are quantum field theories, which describe particles and their interactions in terms of dynamical fluctuations of fields. As opposed to one type of photon which is responsible for the electromagnetic interaction, there are eight types, called colors, of gluons responsible for the strong interactions.  Each flavor of quark (up, down, strange, {\em etc.}) comes in three colors, or more generally in superpositions of the three colors.  Other than by analogy, the term ``color'' when referred to quarks and gluons is a label that is not related to the wavelength of light or to any other optical characteristic.  Color in this context is a separate degree of freedom related  to the strong interactions, not electromagnetism.

The relativistic description of electromagnetism is described in terms of a four-vector field $A^\mu(x)$, $\mu\in\left\{0,1,2,3\right\}$, which satisfies  Maxwell's equations,
\begin{eqnarray}
\frac{\partial}{\partial x^\mu}F^{\mu\nu}&=&  J^\nu, \label{eq:Maxwell1} \\
\epsilon^{\mu\nu\lambda\sigma}\partial_\nu F_{\lambda\sigma}&=&0, \end{eqnarray}
where the Levi-Civita tensor $\epsilon^{\mu\nu\lambda\sigma}$ is completely antisymmetric
in exchange of its indices and $\epsilon^{0123}=1$; $J^\nu$ is the electromagnetic current; and the field strength tensor is given by
\begin{equation}
F_{\mu\nu}=\frac{\partial}{\partial x^\mu}A_\nu-\frac{\partial}{\partial x^\nu}A_\mu. \end{equation}
In special relativity, the distinction between tensors with upper and lower indices depends on the Minkowski metric $\eta_{\mu\nu}$, which is diagonal with $\eta_{00}=1$ and $\eta_{ij}=-\delta_{ij}$ if $i,j\neq0$, where the Kronecker delta $\delta_{ij}=1$ if $i=j$ and $\delta_{ij}=0$ otherwise.  
The components of $\eta^{\mu\nu}$ are the same as those of $\eta_{\mu\nu}$.  Indices are raised and lowered with the Minkowski metric, so for example, $F^{\mu\nu}=\eta^{\mu\alpha}\eta^{\nu\beta}F_{\alpha\beta}$, where a sum over repeated Lorentz indices is always assumed.

The half of Maxwell's equations given by Eq.~(\ref{eq:Maxwell1}) follow from a Lagrangian density,
\begin{equation}
L_{{\rm EM}}=-\frac{1}{4}F_{\mu\nu}F^{\mu\nu}+ A_\mu J^\mu.  \label{eq:LQED}\end{equation}
In QCD all of this is the same, except that $A_\mu$ is replaced by the gluon fields $A_\mu^a$, $a\in\left\{1,\dots,8\right\}$, and $J^\mu$ is replaced by the color current $J^{a\,\mu}$: \begin{equation}
 L_{{\rm QCD}}=\sum_{a=1}^8\left[-\frac{1}{4}F_{\mu\nu}^aF^{a\,\mu\nu}+A_\mu^a J^{a\,\mu}\right].\label{eq:LQCD}\end{equation}
The gluon field strength tensor differs from the electromagnetic field strength tensor in an important way: due to the non-Abelian gauge invariance of QCD the field strength tensor is nonlinear in the gluon fields $A_\mu^a$.  In particular,  \begin{equation}
F_{\mu\nu}^a=\frac{\partial}{\partial x^\mu}A_\nu^a-\frac{\partial}{\partial x^\nu}A_\mu^a-g_s
\sum_{b,c=1}^8 f^{abc}A_\mu^b A_\nu^c, \label{eq:Fmna}\end{equation}
where $f^{abc}$ are a set of numbers known as the structure constants of the group SU(3), and $g_s$ is the coupling constant of the strong interactions, analogous to $e$ in electromagnetism.
In addition to the Lagrangian in Eq.~(\ref{eq:LQCD}) there are kinetic terms for the quark fields, just as there are kinetic terms for electrons in QED, but the interactions are governed by the Lagrangians above.

The historical development of QCD contains many surprising  discoveries, both experimental and theoretical.  As a result of a Nobel Prize-worthy minus sign  which leads to an anti-screening effect  in QCD \cite{asymptoticfreedom}, the strong interactions behave quite differently than electromagnetism.  The eight gluons responsible for the strong interactions couple to matter differently than eight copies of the photon would, and due to the term in Eq.~(\ref{eq:Fmna}) quadratic in the gluon fields, gluons also couple to themselves.
Even if there were eight copies of the photon, and correspondingly eight sets of Maxwell's equations and eight different kinds of electric charges, matter would still give rise to (eight copies of) the standard Coulomb potential and would obey the same type of Lorentz force law for each  of the eight electromagnetisms.  The electric field lines between a static proton and electron  have a typical dipole distribution, and in particular the field lines spread out far from the charged particles.  

The strong interactions are quite different in this respect.  While at distances smaller than around a fermi ($10^{-15}$~m) from a quark, an antiquark would experience a Coulomb-type potential due to the strong interactions, at slightly larger distances the field lines between a quark and an antiquark become confined into a flux tube which binds the two particles. At still larger distances this classical picture of the gluon field breaks down.  A bound quark-antiquark pair (called a meson) would have so much energy if the quark and antiquark were stretched apart more than a couple of fermi that new particles would be created in the process and the flux tube would dissociate into different smaller configurations of the gluon field.  This is what happens when showers of particles are created by high-energy scattering at particle colliders.  Incidentally, the attempt to quantize the gluon flux tube, or ``string,''  led to the original development of string theory in the 1960s.

So, the reason we do not regularly experience the strong interactions as we do electromagnetism and gravity is that the gluons, although they are massless, do not escape the confines of nuclear-sized bound-state hadrons.  Incidentally, a rigorous mathematical proof of the confining
property, and even the mathematical consistency, of QCD is still lacking, and is one of the Millenium Prize Problems posed by the Clay Mathematics Institute in 2000 \cite{MilleniumPrize}.

The hadrons are messy objects when probed at short distances where the strong interactions are relevant, but at longer distances are characterized by only a few quantum numbers.  Hadrons are specified by their mass, spin, electric charge, and transformation properties under a few  discrete and approximate symmetries, such as charge conjugation, parity, and isospin.  Typically, for every hadron with a particular set of quantum numbers, there are others that are identical except for their masses and decay rates.  These hadronic resonances form towers of states known as radial excitations.  For example, the rho meson $\rho(770)$ has spin 1, is an isospin triplet, is odd under both parity and charge conjugation, and has a mass of around 776 MeV/c$^2$ \cite{pdg}.  The $\rho(1450)$ appears to be  identical, except that it has a mass of around 1450 MeV/$c^2$.  Similarly for the $\rho(1700)$, the (tentatively identified) $\rho(1900)$, and the $\rho(2150)$.  Other resonances follow a similar pattern, but with different masses and higher spin.  Holographic QCD reproduces  qualitative, and with limited but surprising accuracy also quantitative, features of the bound states of light quarks in QCD.  In order to understand the extent to which we might expect a higher-dimensional model to accurately describe lower-dimensional physics we need to understand the reason that holographic dualities like the AdS/CFT correspondence are possible.

\section{Black Holes and Holography}
In known holographic dualities, the higher-dimensional version of the theory always contains gravity, and the failure of the intuition that fewer dimensions implies fewer degrees of freedom is related to some unusual properties of black holes in general relativity.  A black hole is a spacetime with a horizon that separates causally distinct regions of the spacetime.  The Schwarzschild black hole is such a spacetime that arises as a solution to Einstein's equations in vacuum, {\em i.e.} with no matter.  Classically, an observer inside the horizon of the Schwarzschild black hole cannot communicate with an observer outside the horizon.  (Quantum mechanically, the situation is less clear \cite{Hawking:2014tga}.)  The size of the horizon is given by the Schwarzschild radius, $R_S$, which depends on the black hole mass $M$ via \begin{equation}
R_S=\frac{2GM}{c^2}, \end{equation}
where $G$ is Newton's gravitational constant and $c$ is the speed of light.  The area of the black hole horizon is defined as $A_H=4\pi R_S^2$.

The seminal work of Bardeen, Carter, Bekenstein, and Hawking in the 1970s \cite{BlackHoleThermo} demonstrated that black holes have a thermodynamic description, but as such black holes are quite unusual.  Hawking discovered that quantum fields in the background of a black hole radiate with a temperature dependent on the mass of the black hole \cite{Hawkingrad}.  
The temperature of Hawking radiation, $T_H$, varies with the mass $M$ as \cite{Hawking:2014tga}, \begin{equation}
T_H=\frac{\hbar c^3}{8\pi GMk_B}. \end{equation}
If the energy of the black hole is $E=Mc^2$, the temperature is $T_H$ and the entropy is $S$, then from the first law of thermodynamics, \begin{equation}
d(Mc^2)=T_H dS=\frac{\hbar c^3}{8\pi GMk_B}dS. \end{equation}
Integrating, with the condition that the entropy vanishes if $M=0$, gives \begin{equation}
S=\frac{4\pi Gk_BM^2}{\hbar c}=\frac{c^3k_B}{4G\hbar} A_H. \label{eq:SAH} \end{equation}

Entropy is a measure of the number of ways that a system can be reconfigured while not changing its macroscopic features.  The maximum entropy of a system is related to the number of degrees of freedom of the system.  Under normal circumstances  entropy is an extensive quantity: the entropy of a combination of two independent subsystems is the sum of the entropies of those subsystems. As a result, entropy typically varies with the volume of a system.  In the presence of a black hole, Eq. (\ref{eq:SAH}) implies that entropy is not an extensive quantity: entropy does not vary as a volume, but rather as an area.  In 1993, based on this observation, Gerard 't Hooft \cite{tHooft} suggested that in any quantum mechanical system with gravity, physics is fundamentally described by degrees of freedom in one spatial dimension fewer than in the absence of gravity.    Leonard Susskind formalized this holographic principle and suggested how it might be implemented \cite{Susskind}.
The AdS/CFT correspondence is an explicit realization of the holographic principle and is the motivation for holographic QCD.

\section{Holographic QCD}
One of the earliest quantitative comparisons of a model based on the AdS/CFT correspondence to QCD was the work of Cs\'aki, Ooguri, Oz and Terning in 1998 \cite{Glueballs}, which considered the spectrum of scalar bound states of gluons known as glueballs.  Like other hadronic states, the glueballs have a spectrum of massive resonances, and Ref.~\cite{Glueballs} compared AdS/CFT predictions for the spectrum of some of these states to estimates from lattice QCD.  

The existence of towers of hadronic states with identical quantum numbers but different masses is reminiscent of the situation in quantum-mechanical systems with bound states.  
For example, a particle in a one-dimensional box has quantized energies.
The orbitals of the electron in the hydrogen atom are labeled by the principal quantum number $n$ and the angular momentum quantum numbers $(l,m,m_s)$.  Fixing the $(l,m,m_s)$ quantum numbers does not uniquely specify the state; for each consistent set of $(l,m,m_s)$ there is a tower of states with different energies labeled by an integer $n> l$.   The bound states in QCD with given angular momentum and other quantum numbers similarly form towers of states known as radial excitations with otherwise identical quantum numbers.

\subsection{Kaluza-Klein Modes in a Section of Flat Spacetime}

Suppose that the rho meson could propagate in an extra dimension of finite extent.  The quantum mechanical description of this rho meson would be  that of a relativistic particle  free to move in three spatial dimensions but confined to an interval in a fourth dimension.  By tradition we  label the coordinate of the extra dimension $z$.  The fourth spatial dimension creates an infinite-square-well type box, say of size $L$, for the rho meson.   Consider the momentum that this extra-dimension-dwelling rho meson might have.
The eigenstates of the momentum operator $P=-i\hbar\left(\nabla_3,\frac{\partial}{\partial z}\right)$, have wavefunctions  \begin{equation}
\psi_{\mathbf{k},k_z}=e^{i\left(\mathbf{k}\cdot\mathbf{x}+k_z z\right)}, \end{equation}
where the wavevector components $k_1$, $k_2$ and $k_3$ can take any values, but due to the boundary conditions, which we will assume for now are $\psi(z=0)=0 $ and $\psi(z=L)=0$, the 
allowed states will be superspositions  for which the wavevector
component $k_z$ only takes discrete values labeled by an integer $n$,\begin{equation}
k_{zn}=\frac{n\pi}{L}. \end{equation}
The corresponding momenta are \begin{equation}
\left(\mathbf{p},p_{zn}\right)=\left(\hbar\mathbf{k},\frac{\hbar n \pi}{L}\right). \end{equation}
The energy eigenstates which satisfy the hard-wall boundary conditions are actually superpositions of momentum eigenstates with $z$-component  $\pm p_{zn}$, just as for the nonrelativistic quantum-mechanical particle-in-a-box.
If the rest mass of the extra-dimensional rho meson is $m_0$, then in a relativistic theory this momentum would contribute to the energy of the rho meson via the relation, \begin{eqnarray}
E^2&=&(\mathbf{p}^2+p_{zn}^2)c^2+m_0^2 c^4 \nonumber \\
&=&\hbar^2\mathbf{k}^2c^2+\left(\frac{\hbar^2n^2\pi^2}{L^2c^2}+m_0^2\right)c^4.
\label{eq:relEp}\end{eqnarray}
The terms on the right-hand-side of  Eq.~(\ref{eq:relEp}) have been collected in a suggestive fashion. From the perspective of the physicist who insists that physics be described in terms of a four-dimensional spacetime, the extra-dimensional rho meson is replaced with a tower of rho mesons with different masses labeled by the integer $n$, via \begin{equation}
m_n^2= \frac{\hbar^2n^2\pi^2}{L^2c^2}+m_0^2. \label{eq:mnFlat}\end{equation}
These modes are often referred to as Kaluza-Klein modes, and the analysis above is
the basis of modern Kaluza-Klein theories: if fields (or particles) propagate in one or more compact extra dimensions, then from a lower-dimensional perspective each field becomes a tower of fields (or particles) with otherwise identical quantum numbers, but different masses.  In general, the extra-dimensional spacetime may be described by a less trivial geometry, in which case the spectrum would differ from the particle-in-a-box spectrum above.

This brings us back to holography and QCD. We are motivated to consider the possibility that rho mesons, and other hadrons, might be modeled as Kaluza-Klein modes of particles that propagate in an extra dimension with an appropriate spacetime geometry.  Such a model would be holographic in the sense that it relates QCD in 3+1 spacetime dimensions to an alternative description in more spacetime dimensions.  We will be using the holographic analogy in reverse: the higher-dimensional description will be used to study the properties of the lower-dimensional system.  The reasoning that led to the early conjectures of holographic dualities relied on the properties of gravity, so if there is a precise sense in which QCD is dual to a higher-dimensional theory, we should expect the higher-dimensional picture to include gravity.  Conveniently, gravity is described by a spin-two field in general relativity, and in QCD there are spin-2 bound states  called tensor mesons, which would be described by the higher-dimensional gravitons via this type of duality.  By assuming a specific extra-dimensional spacetime background the spectrum of tensor meson masses may be predicted, as was done in Ref.~\cite{Katz:2005ir}.

The rho meson is a spin-1 particle, which is the type of particle that arises from a vector field like the electromagnetic field.  (A massive photon would have helicity $\pm\hbar$ or 0.) The ``AdS'' in the AdS/CFT correspondence stands for anti-de Sitter space.  This is the maximally symmetric spacetime with constant negative curvature.  In string theory it arises as the description of a region of space around a stack of a large number of $D$-branes, which are dynamical manifolds  on which open strings can end \cite{D-branes,Gauntlett:1998ht}. Anti-de Sitter space also arises as a solution to Einstein's equations in the presence of a negative cosmological constant but in the absence of any kind of matter.  For comparison, the accelerated expansion of the universe might be due to a positive cosmological constant, which in the absence of matter would generate another maximally symmetric spacetime known as de Sitter space.   

Motivated by the AdS/CFT correspondence, phenomenological bottom-up models of holographic QCD typically assume that the extra-dimensional fields propagate in a portion of anti-de Sitter space.  In addition to simplicity, there is some phenomenological motivation for this choice.  At high energies, QCD is asymptotically free and effectively has no dimensionful scale governing interactions.  This scale invariance is a property of conformal field theories (CFT), which is the ``CFT"  in the AdS/CFT correspondence.  Furthermore, it has been suggested that QCD is conformal also at very low energies \cite{conformal}, so the AdS geometry seems like an appropriate starting point for  model-building purposes.

In Cartesian coordinates, infinitesimal lengths in $d$-dimensional Euclidean space satisfy the Pythagorean theorem:
\begin{equation}
dl^2=\sum_{i=1}^d dx_i^2. \end{equation}
Lengths are invariant under rotations of the coordinates in Euclidean geometry.  Similarly, the infinitessimal  interval in Special Relativity, \begin{equation}
ds^2=c^2 dt^2-\sum_{i=1}^d dx_i^2, \label{eq:Minkds}\end{equation}
is invariant under Lorentz transformations.  The similarity between this notion of an invariant in Special Relativity and the invariant length of Euclidean geometry led Poincar\'e and Minkowski \cite{Minkowski} to propose a geometric interpretation of spacetime, which was later generalized by Einstein \cite{Einstein:1916vd} in his geometric theory of gravitation, general relativity.  
In a non-Euclidean geometry there is no coordinate system in which infinitesimal lengths  satisfy the Pythagorean theorem, or in the case of spacetime Eq.~(\ref{eq:Minkds}).
A generic spacetime may be specified by the metric $g_{MN}$, such that the Pythagorean theorem is replaced by \begin{equation}
ds^2=\sum_{M,N=0}^d g_{MN} dx^M dx^N, \end{equation}
where we identify $x^0$ with a time coordinate, and $x^M$ with the spatial coordinates. Except in Cartesian coordinates in  Euclidean or Minkowski space, the metric $g_{MN}$  depends on the coordinates $x^M$.  This is a good time to simplify our equations following relativistic conventions: We use units such that the  speed of light  $c=1$; and we use Einstein's summation convention, namely that repeated indices in an equation are summed over as in Eq.~(\ref{eq:LQED}) and (\ref{eq:LQCD}).

The metric of special relativity is the Minkowski tensor $\eta_{MN}$,  The matrix inverse of the metric $g_{MN}$, considered as a matrix with row index $M$ and column index $N$, is denoted with upper indices, $g^{MN}$.  Indices on other vectors and tensors are raised with $g^{MN}$ and lowered with $g_{MN}$, so for example, $A^M=g^{MN}A_N$, and $A_M=g_{MN}A^N$.

Rather than consider the particle-like momentum and energy of Kaluza-Klein modes that might be interpreted as rho mesons, we may instead consider a wavelike description.  The rho mesons are vector mesons, and in this description would be described by vector fields in an extra dimension, and satisfy Maxwell's equations in the higher-dimensional spacetime.  In a convenient choice of gauge with $A_z=0$, components of the electromagnetic 4-vector $A_\mu$ satisfy the wave equation:
\begin{equation}
-\frac{1}{c^2}\frac{\partial^2 A_\mu}{\partial t^2}+\nabla_3^{\, 2} A_\mu+\frac{\partial^2 A_\mu}{\partial z^2}=0,\end{equation}
where $\nabla_3^{\,2}$ is the three-dimensional Laplacian, $z$ is the coordinates of the extra dimension, which exists over the region $z\in(0,L)$, and $c$ is the speed of light, which we again set  to $c=1$.  The extra dimension here is of finite extent, so we are describing a higher-dimensional waveguide.  The wave equation is separable into dependence on $z$ and on the remaining 3+1 spacetime coordinates; the discrete tower of solutions to this equation are Kaluza-Klein modes.  In particular, we can find solutions of the form $A_\mu(x^\nu,z)=a_\mu(x)\psi(z)$.  
The wave equation separates: \begin{equation}
\eta^{\mu\lambda}\frac{\partial^2 a_\nu(x)}{\partial x^\mu \partial x^\lambda}=-m^2 a_\nu(x),
\label{eq:a1}\end{equation}
\begin{equation}
\frac{d^2\psi(z)}{d z^2}=-m^2\psi(z). \label{eq:psi1}
\end{equation}
Boundary conditions on $\psi(z)$ at the walls of the waveguide $z=0$ and $z=L$ determine the eigenvalues $m$, and consequently the spectrum of Kaluza-Klein modes.  For example, the boundary conditions $F_{\mu z}(0)=F_{\mu z}(L)=0$ in this gauge are equivalent to the  conditions $\psi'(0)=\psi'(L)=0$, in which case the solutions to Eq.~(\ref{eq:psi1}) are \begin{equation}
\psi(z)=A \cos\left(m z\right),\end{equation}
where $A$ is an arbitrary constant.  The eigenvalue $m$ satisfies $\sin(mL)=0$ and the allowed values of $m$ are $m_n=n\pi/L$, for integer $n$.  The analogy with Eq.~(\ref{eq:mnFlat}) should be clear, and indeed, the value of $m$ in the Proca generalization of Maxwell's equations, Eq.~(\ref{eq:a1}), is interpreted as the mass of the vector field.

This was a warm-up to  holographic QCD, in which the higher-dimensional spacetime has a nontrivial geometry, the prototypical example being a section of anti-de Sitter space.

\subsection{Kaluza-Klein Modes in a Section of Anti-de Sitter Spacetime}
\begin{figure}[h]
\includegraphics[width=2.5in]{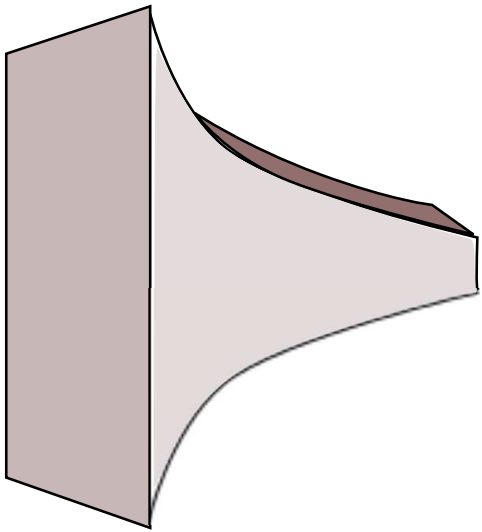}
\centering
\caption{Proper lengths are rescaled as one moves around in anti-de Sitter space.}
\label{fig:AdS}
\end{figure}

In the anti-de Sitter spacetime there is no coordinate system in which lengths follow the Pythagorean theorem familiar from Euclidean geometry.   Instead, infinitessimal intervals in anti-de Sitter space are given by, 
\begin{equation}
ds^2=\frac{R^2}{z^2}\left(dt^2-dx_1^2-dx_2^2-dx_3^2-dz^2\right). 
\end{equation}
Equivalently, anti-de Sitter space may be described by the metric \begin{equation}
g_{MN}=\frac{R^2}{z^2}\eta_{MN}, \label{eq:AdS}\end{equation}
where $R$ is the AdS scale.   The factor $R^2/z^2$ multiplies the metric of flat spacetime, so we say that anti-de Sitter space is conformally flat, which means that it is like Euclidean space except that infinitessimal lengths are rescaled by a position-dependent factor.  Infinitessimal time differences are rescaled the same way.
We can think of the factor of $R^2/z^2$ as describing how lengths are rescaled as one moves along the ``extra dimension'' $z$, but the choice of $z$ as a special coordinate is a consequence of our coordinate system and not of any one direction as being special in an absolute sense.  In these coordinates a stick of coordinate-length one meter lying along the  three spatial dimensions $x_1,\,x_2$ and $x_3$ has four times the proper length of a stick of the same coordinate size when moved over in the extra dimension to a point with twice the value of the $z$-coordinate.  This warping of the spacetime is  depicted in Fig.~\ref{fig:AdS}, and the relation between the extra dimension and the length scale of interest is made explicit in an approach known as light-front holographic QCD  \cite{BdT,Brodsky:2004tq}.
The slice of the spacetime with $z=0$ is a boundary of the spacetime.  To describe a discrete tower of states, we will consider Maxwell's theory in the portion of anti-de Sitter space between $z=0$ and $z=z_0$ for some $z_0$.  The hard wall at $z=z_0$ creates a waveguide, and results in a discrete spectrum of fields analogous to the discrete tower of states of the quantum-mechanical particle in a box.

Holographic QCD models are similar to electromagnetism in curved spacetime.  The equations of motion, which are Maxwell's equations in anti-de Sitter space, are given in Eqs.~(\ref{eq:Az0eq1}) and (\ref{eq:Az0eq2}), and the separable solutions to those equations in a convenient gauge choice satisfy Eqs.~(\ref{eq:a}) and (\ref{eq:psi}). The following discussion explains the derivation of those equations for readers adept at field theory in curved spacetime, but other readers may want to skip ahead to the equations of motion.  

The equations of motion for the electromagnetic field in anti-de Sitter space follow from the Lagrangian \begin{equation}
L_{{\rm EM}}^{({\rm AdS})}=\sqrt{|g|}\left[-\frac{1}{4}F_{MN}F^{MN}+ A_M J^M\right], \label{eq:LEMAdS}
\end{equation}
where $g$ is the determinant of the metric $g_{MN}$.  The various factors of the metric and its determinant in the Lagrangian  make the formulation of the theory independent of coordinate system, which is the basic principle underlying general relativity.  The equations of motion, {\em i.e.} (half of) Maxwell's equations in curved spacetime, are the Euler-Lagrange equations for the Lagrangian $L_{{\rm EM}}^{({\rm AdS})}$:
\begin{equation}
\frac{\partial}{\partial x^M}\left(\sqrt{|g|}F^{MN}\right)=\sqrt{|g|}J^N. \end{equation}

From here on we consider the electromagnetic field in the absence of sources, {\em i.e.} $J^N=0$.
With the AdS metric of Eq.~(\ref{eq:AdS}) in 4+1 dimensions, $g=1/z^{10}$.  With the convention that Greek indices $\mu$, $\nu$, {\em etc.} take values in $\{0,1,2,3\}$, the equations of motion become, \begin{equation}
\frac{\partial}{\partial z}\left(\frac{R^5}{z^5}F^{zN}\right)+\frac{R^5}{z^5}\frac{\partial}{\partial x^\mu}F^{\mu N}=0. \label{eq:EOMwR}\end{equation}
The AdS scale $R$ factors out of Eq.~(\ref{eq:EOMwR}) and subsequent analysis, so we henceforth set $R=1$.
Lowering indices introduces additional factors of $z^2$ for each index lowered, from the inverse metric $g^{MN}=z^2 \eta^{MN}$:
\begin{equation}
\frac{\partial}{\partial z}\left(\frac{1}{z}F_{zN}\right)+\frac{1}{z}\frac{\partial}{\partial x^\mu}F_{\lambda N}\eta^{\mu\lambda}=0. \end{equation}
This is the form of the equation of motion we will work with.

\begin{figure}[h]
\includegraphics[width=3in]{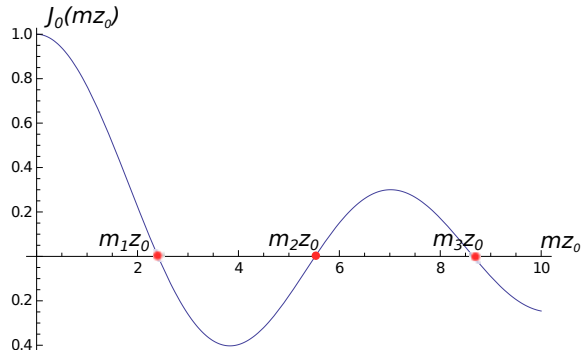}
\centering
\caption{Vector meson masses are determined by zeroes of the Bessel function $J_0(mz_0)$ in a hard-wall model of holographic QCD, where $z_0$ is the location of the hard wall.}
\label{fig:J0}
\end{figure}

Just as in flat spacetime, the electromagnetic action and equations of motion are  invariant under local transformations known as gauge transformations of the form
\begin{equation}
A_M\rightarrow A_M+\frac{\partial f}{\partial x^M}, \end{equation}
where $f$ is an arbitrary function of the spacetime coordinates.  Unlike global symmetries that do not depend on the location in spacetime, gauge invariances indicate a redundancy in the description of the physics.
It is often convenient to choose a gauge, such as Lorenz gauge $\partial_M A^M=0$, to simplify the form of Maxwell's equations in terms of $A^M$.  We will choose the axial  gauge $A_z=0$, which once fixed still allows gauge transformations with functions $f$ independent of the extra-dimension coordinate $z$.  
The curved-space Maxwell equations in $A_z=0$ gauge take the form,
\begin{equation}
\frac{\partial}{\partial z}\left(\frac{1}{z}\frac{\partial A_\nu}{\partial z}\right)+\frac{1}{z}\frac{\partial}{\partial x^\mu}F_{\lambda \nu}\eta^{\mu\lambda}=0,
\label{eq:Az0eq1}\end{equation}
\begin{equation}
-\frac{1}{z}\frac{\partial}{\partial x^\mu}\left(\frac{\partial A_\lambda}{\partial z}\right)
\eta^{\mu\lambda}=0. \label{eq:Az0eq2}
\end{equation}
Interchanging the order of the mixed partial derivatives in Eq.~(\ref{eq:Az0eq2}), it follows that $\eta^{\mu\lambda}\partial A_\lambda/\partial x^\mu$ is independent of the extra-dimensional coordinate $z$.  As a result, we can use the residual gauge freedom to set  $\eta^{\mu\lambda}\partial A_\lambda/\partial x^\mu=0$.  Then Eq.~(\ref{eq:Az0eq1}) simplifies:
\begin{equation}
\frac{\partial}{\partial z}\left(\frac{1}{z}\frac{\partial A_\nu}{\partial z}\right)+\frac{1}{z}\frac{\partial}{\partial x^\mu}\left(\frac{\partial A_\nu}{\partial x^\lambda}\right)\eta^{\mu\lambda}=0.
\label{eq:AEOM}\end{equation}
In this gauge, the equation of motion for each component of $A_\nu$ is independent.  The Kaluza-Klein modes, which would be interpreted as a tower of massive 3+1 dimensional electromagnetic fields, are solutions to Eq.~(\ref{eq:AEOM}) in which the $z$-dependence has been separated from the $x^\mu$-dependence:
\begin{equation}
A_\mu(x^\nu,z)=a_\mu(x)\psi(z).\end{equation}
Eq.~(\ref{eq:AEOM}) separates: \begin{equation}
\eta^{\mu\lambda}\frac{\partial^2 a_\nu}{\partial x^\mu \partial x^\lambda}=-m^2 a_\nu,
\label{eq:a}\end{equation}
\begin{equation}
\frac{d}{d z}\left(\frac{1}{z}\frac{d\psi}{dz}\right)=-\frac{m^2}{z}\psi. \label{eq:psi}
\end{equation}
We need boundary conditions at $z=0$ and $z=z_0$.  At $z=0$ we choose Dirichlet boundary conditions $\psi(0)=0$, and at $z=z_0$ we choose $F_{\mu z}=0$. For the Kaluza-Klein modes the boundary conditions become $\psi(0)=0$ and  $d \psi/d z=0$ at $z=z_0$.  The solutions to Eq.~(\ref{eq:psi}) with these boundary conditions determine the eigenvalues for $m^2$.  From Eq.~(\ref{eq:a}) the lower-dimensional interpretation of these eigenvalues is the spectrum of masses (squared) of the tower of electromagnetic fields described by $a_\mu(x)$, or of the corresponding photons.
Eq.~(\ref{eq:psi}) has solutions in terms of Bessel functions, and the boundary conditions determine the eigenvalues \cite{Erlich:2005qh,Da Rold:2005zs}, \begin{equation}
J_0(m_n z_0)=0,  \label{eq:rhomasses}\end{equation}
where we have labeled the discrete tower of eigenvalues by an integer $n$, as indicated for the first three modes in Fig~\ref{fig:J0}.

The exercise above was framed in terms of electrodynamics in order to introduce some of the necessary formalism, but holographic QCD addresses the properties of hadrons arising from the strong interactions, not electrodynamics.  The rho mesons are described by triplets of fields (they transform in the spin-1 representation of the SU(2) isospin symmetry), but otherwise they are similar to massive photons. In particular, they are spin-1 particles, one of which is electrically neutral.  
In order to describe the dynamics of rho mesons holographically, we need to choose a spacetime and a Lagrangian for the appropriate higher-dimensional fields.  For simplicity we consider the same slice of anti-de Sitter space as above.  One lesson from the known examples of holographic dualities is that symmetries of the lower-dimensional description of the theory become gauge invariances in the higher-dimensional description.  The SU(2) isospin symmetry, which is the approximate symmetry of QCD which neglects the difference between the up and down quark masses, suggests that the holographic model should be an SU(2) gauge theory.  The rho mesons would then be described by the Kaluza-Klein modes of the SU(2) gauge fields.  

The linearized equations of motion for the three SU(2) gauge fields are each identical to the equations for the electromagnetic field described above, so the spectrum of masses $m_n$ determined by Eq.~(\ref{eq:rhomasses}) in this model are interpreted as the spectrum of rho mesons.  The lightest rho meson, with mass around $m_\rho=770$~MeV, may be used to fix $z_{0}$, in which case $z_{IR}=1/(320 {\rm MeV})$.  Then, we could use Eq.~(\ref{eq:rhomasses}) to predict the spectrum of heavier rho resonances, which depend on zeros of $J_0$.  With $m_1=770$~MeV, we find $m_2=1770$~MeV and $m_3=2770$~MeV for the next few resonances.  This spectrum does not agree well with the measured values $m_2=1450$~MeV and $m_3=1700$ MeV.  Hence, this toy model might qualitatively capture some of the properties of the tower of rho resonances, but fails to  predict the spectrum of these resonances.  

Before giving up on this model, another thing that the model predicts is the rho meson decay constant, $F_\rho$, which can be interpreted as the mixing between the neutral rho meson and the photon. The decay constant may be determined from the effective Lagrangian obtained by replacing the higher-dimensional gauge field in Eq.~(\ref{eq:LEMAdS}) with a decomposition in Kaluza-Klein modes, after adding to that decomposition a zero-mode, {\em i.e.} a solution to the same equation of motion with eigenvalue $m^2=0$.  The zero-mode represents the photon, and the mixing between the photon and the rho meson can be read off of the effective action as the term proportional to $F^{\mu\nu}_{{\rm photon}}F_{{\rm rho}\,\mu\nu}$ in the decomposition of the action in terms of Kaluza-Klein modes \cite{Sakai-Sugimoto}.  There is an alternative approach motivated by the AdS/CFT correspondence which makes use of the behavior of the solutions to the equations of motion near the AdS boundary at $z=0$ to determine the decay constant from the correlation function of a product of isospin currents \cite{Erlich:2005qh,Da Rold:2005zs}, but the results of these two approaches are identical.  The predicted decay constant depends on $z_0$ and the gauge coupling.  With the model parameter $z_0$ determined from $m_\rho=770$ MeV, and the gauge coupling fixed by matching to the short-distance behavior of the isospin current-current correlator \cite{Erlich:2005qh,Da Rold:2005zs}, the result is $F_\rho=(330\ \rm{MeV})^2$ \cite{Erlich:2005qh}, which is to be compared to the experimental value of $(345\pm8)^2$ ${\rm MeV}^2$ \cite{pdg}.  This is a prediction for just one observable, but in the absence of a useful statistical measure of the success of this crude model we might just say that the agreement between the predicted and measured values of $F_\rho$ is relatively good.

The failure of the model to predict the properties of the heavier resonances is easily understood.  The mass eigenvalues $m_n^2$ determined by this model grow with resonance number $n$ roughly like the  energy eigenvalues for the nonrelativistic particle in a box: $m_n^2\sim n^2$.  This result is relatively generic, and in particular does not rely on the precise form of the spacetime, which we chose somewhat arbitrarily based on the spacetime of the prototypical AdS/CFT correspondence. However, the spectrum of hadrons roughly follows the Regge behavior $m_n^2\sim n$.  This is a problem for typical holographic models of QCD \cite{Schreiber:2004ie,Shifman:2005zn}.  However, by introducing a new scalar field, called the dilaton, into the model, the spectrum of mesons predicted by holographic QCD can be made to agree with the Regge behavior \cite{Karch:2006pv}.  Models of this type are called {\em soft-wall} models because the spacetime does not rely on  hard-wall boundary conditions  in order to obtain a discrete spectrum of states.  In fact, the introduction of a dilaton scalar into the model is well motivated by the AdS/CFT correspondence, in which modes of the dilaton correspond to scalar glueballs.

Another way to understand the failure of holographic QCD at high energies is to compare the model to the holographic duals that arise in string theory.  In particular, in order to decouple stringy physics so that classical calculations like the one performed above are sufficient, the gauge theory described by the duality must have the analogy of an infinitely large number of colors \cite{Maldacena:1997re}.  The quarks in QCD have three colors, so stringy corrections are generally not negligible.  The Regge spectrum can be understood by modeling the mesons in terms of fluctuations of a flux tube of gluons binding quarks and antiquarks, so it is not surprising that we miss that behavior without somehow modifying a simplistic model like the hard-wall model.  The hard-wall model successfully reproduces low-energy hadronic observables where stringy corrections are less important, but at energies significantly higher than the rho meson mass we should not expect holographic QCD to work.  Other unphysical aspects of the large-$N$ limit of QCD that persist in holographic QCD calculations include incorrect jet structure of high-energy scattering events \cite{Csaki:2008dt}; and  the vanishing of resonance widths, although one can self-consistently introduce resonance widths by taking into account nonvanishing couplings between the hadrons in holographic QCD.

Having understood the problem with the spectrum of heavier resonances in hard-wall holographic QCD models, we might still wonder how to generalize these models to describe hadrons besides the rho meson, where we will intend these models to describe the hadrons only at low-enough energy that stringy physics is unimportant.  One approach is to include additional approximate symmetries of QCD into the model, such as the chiral symmetry that arises if the masses of the up and down quarks are neglected altogether.  Due to the chiral symmetry there is an additional SU(2) gauge invariance in the resulting holographic model.  An interesting aspect of QCD is that, even if the chiral symmetry were exact, the  symmetry is spontaneously broken by the vacuum expectation value of an operators known as the chiral condensate.  In other words, the ground state of QCD does not manifest the underlying symmetry of the theory, just as a crystal does not manifest the translation invariance of the dynamics underlying the atoms in the crystal, and the ground state of a ferromagnet at low temperatures does not respect rotation invariance.  Whenever there is a spontaneously broken continuous global symmetry in a Lorentz-invariant system, Goldstone's theorem implies that there is a massless, spinless particle in the spectrum.  In the case of the broken chiral symmetry the Goldstone modes are particles called pions.  The chiral symmetry in QCD is only approximate, so the pions are not massless, but at around 140 MeV are significantly lighter than all of the other hadronic bound states.  

In the higher-dimensional holographic model the chiral symmetry is a gauge invariance, so its breaking involves the breaking of a gauge invariance.  In the Standard Model, a nonvanishing vacuum expectation value for the Higgs field is responsible for breaking of the gauge invariance of the electroweak interactions.  By analogy, the models of Ref.~\cite{Erlich:2005qh,Da Rold:2005zs} include  scalar field which plays the role of the chiral condensate and has a background that spontaneously breaks the gauged chiral symmetry.  Those models predict not only the mass and decay constant of the rho meson, but also similar properties of the axial-vector meson (called $a_1$), and the pions.  A fit of the model of Ref.~\cite{Erlich:2005qh} is reproduced in Table~\ref{tab:EKSS}.  Although that model is crude (it has only three free parameters, analogous to the chiral condensate, quark mass, and $z_0$), the quantitative predictions for these observables are surprisingly accurate.

\begin{table}[ht]
\centering
\begin{tabular}{|c|c|c|}
\hline
Observable&Measured&Model Fit \\
&(Central Value-MeV)&(MeV) \\
\hline
$m_\pi$&139.6&141 \\
$m_\rho$&776&832\\
$m_{a_1}$&1230&1220 \\
$f_\pi$&92.4&84.0 \\
$F_\rho^{1/2}$&345&353 \\
$F_{a_1}^{1/2}$&433&440 \\
$g_{\rho\pi\pi}$&6.03&5.29 \\
\hline
\end{tabular}
\caption{Fit of a three-parameter hard-wall model for masses, decay constants, and  $\rho\pi\pi$ coupling, from Ref.~\cite{Erlich:2005qh}.}
\label{tab:EKSS}
\end{table}
Some of the other observables that have been calculated in this and similar models include the spectrum of tensor mesons: the ratio of the $f_2$ mass to the rho mass is predicted to be 
$m_{f_2}/m_\rho=1.59$, compared with the experimental central value of 1.64 \cite{Katz:2005ir}; the charge radius of the rho meson $r^2_C=0.53$ fm$^2$ \cite{Grigoryan:2007vg}; and the gravitational radius of the rho meson $r^2_C=0.21$ fm$^2$ \cite{Abidin:2008ku}.  In the same model, appropriate backgrounds for the higher-dimensional gauge fields can be chosen so as to correspond to  nonvanishing  chemical potentials, and information can be obtained about the phase structure of holographic QCD, for example the pion condensate phase at large isospin chemical potential  \cite{pion-condensate}.

An independent approach to modeling hadronic physics holographically is based on the observation  that the equations of motion for fields in anti-de Sitter space resemble the equations for hadronic wavefunctions in a Hamiltonian formalism known as light-front QCD \cite{BdT,Brodsky:2004tq} that was mentioned earlier.  In the identification between the two formalisms, the extra-dimensional coordinate $z$  maps into a kinematical variable related to the longitudinal momentum fraction carried by partons (quarks) in the hadron and the  separation between the partons.  Given a particular light-front holographic QCD model, the identification of the two approaches also predicts the form of the confining potential in the light-front Hamiltonian \cite{LFHQCD-confinement}, which is required in order to precisely predict hadronic properties in that approach.

\section{Top-Down AdS/QCD}

The hard-wall and soft-wall holographic QCD models are phenomenological bottom-up  models unconstrained by the rules of string theory.  They are motivated by the AdS/CFT correspondence and certain features of QCD, but they are not precise duals of QCD, nor of any known 3+1 dimensional theory.  There are also top-down models  based on D-brane configurations in string theory, which are are thought to be true holographic duals of theories that possess certain features of QCD, and like bottom-up models can be compared with experimentally measured hadronic observables.  Like bottom-up models, top-down holographic QCD models do not precisely describe QCD, and so far they always contain additional non-QCD states that are ignored when comparing with data.   Holographic QCD models are sometimes referred to as AdS/QCD models, particularly in the case of top-down models rooted in string theory.

Type  IIB string theory is a 9+1 dimensional theory of relativistic strings \cite{string-theory}.  This string theory describes closed loops of string propagating in the bulk of the 9+1 dimensions, but open strings may exist if they end on D-branes, which are dynamical manifolds that exist only in certain spacetime dimensions \cite{D-branes}.   Type IIB string theory, which is relevant for the discussion of the prototypical example of the AdS/CFT correspondence, contains only D-branes with odd numbers of spatial dimensions.  At long distances, the strings which end on a  stack of $N$ D3 branes in Type IIB string theory are described by a 3+1 dimensional field theory known as ${\cal N}$=4 U($N$) Supersymmetric Yang-Mills theory.  It is a U($N$) gauge theory, and in that sense is similar to QCD, a SU(3) gauge theory.  However, the fermions in the theory transform in the adjoint representation of the gauge group, like the gluons, which is different from QCD in which the quarks transform in the fundamental representation of SU(3).

The prototypical AdS/CFT correspondence is related to Type IIB string theory.  At distances large compared to the string length, the Type  IIB string theory is described by a gravitational field theory known as Type  IIB supergravity.  In a certain limit of parameters, if $N$ is large then a stack of D3-branes generates a spacetime geometry which approaches a product of AdS$_5\times$S$^5$.  The AdS$_5$ factor is 4+1 dimensional anti-de Sitter space, while the S$^5$ is a 5-dimensional sphere.  The precise statement of Maldacena's conjectured AdS/CFT correspondence relates 3+1 dimensional ${\cal N}$=4 U($N$) Supersymmetric Yang-Mills theory to Type IIB string theory expanded about a spacetime background of AdS$_5\times$S$^5$.  In a particular limit of small string coupling and large $N$ known as the supegravity limit, the AdS/CFT correspondence becomes a relation between the same gauge theory and Type IIB supergravity in the AdS$_5\times$S$_5$ background.  The dictionary between calculations on the two sides of the duality was given by Witten \cite{Witten:1998qj} and independently by Gubser, Klebanov and Polyakov \cite{Gubser:1998bc}.  The key point is that, while the S$^5$ is a compact spacetime, the AdS$_5$ is noncompact, and has one dimension more than the dual gauge theory.  This is why the duality is holographic.

The ${\cal N}$=4 supersymmetric gauge theory has the property of conformal invariance, so in particular it is scale-invariant.  This is different from QCD, which has a built-in scale, for example the mass of the  rho meson.  The conformal invariance maps into the isometries of the AdS$_5$ geometry, {\em i.e.} the coordinate transformations that leave the dependence of infinitesimal lengths on the coordinates invariant.  In hard-wall models the boundary at $z=z_0$ breaks this invariance and determines the rho meson mass \cite{Polchinski:2001tt}, as we saw in the previous section.  

Another difference between the ${\cal N}$=4 theory and QCD is that, while there are fermions in the theory, there are no quarks, {\em i.e.} fields charged under the gauge invariance in precise analogy to the way quarks carry color charge.  Karch and Katz \cite{Karch:2002sh} studied how fields analogous to quarks could be added to the AdS/CFT correspondence by introducing another type of D-brane into the model, namely D7-branes, which span seven spatial dimensions.  Strings which stretch between the D3 and D7-branes fluctuate in just such a way as to describe quark fields (in addition to bosonic ``squarks").  Still, this model differs from QCD in several ways, and in particular it is lacking a description of chiral symmetry breaking, so  it does not accurately describe pion physics.  An approach to introducing chiral symmetry in this model  was introduced in Ref.~\cite{Babington:2003vm}, by varying the AdS background so as to allow solutions for a field charged under the chiral symmetry to obtain a nonvanishing background profile. 

\begin{figure}[t]
\includegraphics[width=6in]{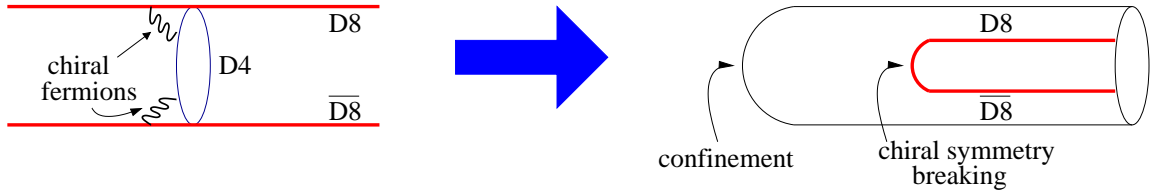}
\centering
\caption{The Sakai-Sugimoto model. D4-branes wrap a circle, and are intersected by D8-branes.  Due to the gravitational backreaction of the D4-branes, the spacetime has a horizon and the D8-branes bend around as shown.}
\label{fig:D4-D8}
\end{figure}

The top-down AdS/QCD model most similar to QCD is the D4-D8 system in Type IIA string theory introduced by Sakai and Sugimoto \cite{Sakai-Sugimoto}.  This model contains D4-branes wrapped on a circle, so that at distances large compared to the circle size the D4-branes appear to have three spatial dimensions.
The D8-branes extend in all but one of the nine spatial dimensions of the string theory, and they intersect the D4-branes on three-dimensional intersections at points along the circle on which the D4-branes are wrapped, as in Fig.~\ref{fig:D4-D8}.  The D8-branes are assumed to come in two stacks, one of which contains anti-D8 branes ($\overline{\rm{D8}}$-branes) which together with the D8-branes break the supersymmetry.  The fermionic fields which propagate on the D4-brane circle have boundary conditions that also break the supersymmetry \cite{Witten:1998zw}.  Like all D-branes, the D4-branes have a tension and act as a gravitational source for spacetime curvature.  The consequence is that the spacetime pinches itself off at a horizon, and in their ground state the D8-branes wrap around the spacetime and connect to the $\overline{\rm{D8}}$-branes as in Fig.~\ref{fig:D4-D8}.  Strings which stretch from the D4-branes to the D8-branes give rise to chiral fermions, so the Sakai-Sugimoto model describes a gauge theory with a continuous chiral symmetry as in QCD.

Whereas originally there is a symmetry associated with separately exchanging the D8-branes among themselves and the $\overline{\rm{D8}}$-branes among themselves, if they connect then there is just an overall symmetry in exchanging the D8 and $\overline{\rm{D8}}$-branes together.  This is just like chiral symmetry breaking: QCD has two independent isospin-like symmetries, one for left-handed and one for right-handed quarks, but only the combination in which the left and right-handed quarks transform together ({\em i.e.} isospin) is manifest in the ground state of the system.  The non-supersymmetric nature of this model makes it more similar to QCD than most top-down models.
The Sakai-Sugimoto model  predicts the light meson spectrum and decay constants, as well as certain hadronic couplings, with similar success to the bottom-up models.  Furthermore, baryons, in analogy to protons and neutrons, arise as solitons in the higher-dimensional spacetime of this model \cite{Sakai-Sugimoto}, which opens new avenues of exploration.

Both bottom-up and top-down holographic QCD models have been extended in various ways, for example to include mesons with strange quarks \cite{Shock:2006qy,Katz:2007tf, Abidin:2009aj}, and to include nonvanishing temperature and density \cite{finiteT,Kovtun:2004de}.  Despite the relatively poor approximation of an enhanced chiral symmetry including the strange quark, the predictions of an extension of the hard-wall model with this symmetry are surprisingly accurate, and a fit of a five-parameter model to fourteen observables is given in Table~\ref{tab:strange} from Ref.~\cite{Abidin:2009aj}.  

\begin{table}[ht]
\centering
\begin{tabular}{|c|c|c|}
\hline
Observable&Measured&Model Fit \\
&(Central Value-MeV)&(MeV) \\
\hline
$m_\pi$&139.6&134 \\
$f_\pi$&92.4&86.6 \\
$m_K$&496&514 \\
$f_K$&113&101 \\
$m_{K_0^*}$&672&697 \\
$f_{K_0^*}$&&36 \\
$m_\rho$&776&789\\
$F_\rho^{1/2}$&345&335 \\
$m_{K^*}$&894&821 \\
$F_{K^*}$& &337 \\
$m_{a_1}$&1230&1270 \\
$F_{a_1}^{1/2}$&433&453 \\
$m_{K_1}$&1272&1402 \\
$F_{K_1^{1/2}}$&&488 \\
\hline
\end{tabular}
\caption{Fit of a five-parameter hard-wall model for meson masses and decay constants, from Ref.~\cite{Abidin:2009aj}.}
\label{tab:strange}
\end{table}

\section{Conclusions}
Holographic QCD successfully reproduces many qualitative features of hadronic physics.   Variations of holographic QCD models  incorporate different aspects of QCD, {\em e.g.} chiral symmetry breaking in the hard-wall model, and the Regge spectrum in the soft-wall model.  The extra-dimensional nature of these models is directly responsible for a number of similarities with QCD and extensions of QCD, particularly in the limit of a large number of colors.  Despite difficulties reproducing aspects of QCD at high energy, these models satisfy sum rules relating correlation functions at high energy to sums over resonances   \cite{Erlich:2005qh,Hong:2004sa}.  They describe vector mesons  in terms of massive vector fields in a manner similar to hidden-local-symmetry models \cite{Bando:1984ej,Son:2003et}, and they explain the observed vector meson dominance in hadronic scattering \cite{Hong:2004sa}.

Holographic QCD models are typically accurate within 10-25\% for QCD observables below around 1.5 GeV, and some of the most accurate predictions are insensitive to model details \cite{Erlich:2008gp}.  Despite their similarities to QCD and their relative success in reproducing experimental data, so far these models have not led to dramatic new insights into confinement or other fundamental aspects of QCD, with possibly one exception: A universal prediction of holographic models is that the ratio of the shear viscosity to the entropy density of a fluid at nonvanishing temperature should have the value $\hbar/4\pi k_B$ \cite{Kovtun:2004de}.  (For suggestions of a holographic description of turbulence, see Ref.~\cite{Eling:2010vr}. The universality of this result is a consequence of the generic properties of the type of black-hole spacetimes that describe the higher-dimensional models, and arguments regarding the reliability of these holographic descriptions suggest that the value $\hbar/4\pi k_B$ is in fact a lower bound for a large class of physical systems \cite{Kovtun:2004de}.  In QCD, a perturbative extrapolation of the same observable for the quark-gluon plasma gives an order of magnitude larger result than the estimate from holographic models  \cite{Csernai:2006zz}. Initial estimates of the viscosity-to-entropy-density ratio in the quark-gluon plasma studied at the Relativistic Heavy Ion Collider appear to agree with the holographic prediction within a factor of a few \cite{Song:2010mg}.  

\section{Acknowledgments}
The author thanks Dylan Albrecht, Graham Kribs, Emanuel Katz, Ian Low, Dam Son,  Misha Stephanov,  Christopher Westenberger and Ron Wilcox for their collaboration on various aspects of holographic QCD.  This work was supported by the NSF under Grant PHY-1068008.


\end{document}